# Unraveling the Chaos-land and its organization in the Rabinovich System

Krishna Pusuluri, Arkady Pikovsky, and Andrey Shilnikov

**Abstract** A suite of analytical and computational techniques based on symbolic representations of simple and complex dynamics, is further developed and employed to unravel the global organization of bi-parametric structures that underlie the emergence of chaos in a simplified resonantly coupled wave triplet system, known as the Rabinovich system. Bi-parametric scans reveal the stunning intricacy and intramural connections between homoclinic and heteroclinic connections, and codimension-2 Bykov T-points and saddle structures, which are the prime organizing centers of complexity of the bifurcation unfolding of the given system. This suite includes Deterministic Chaos Prospector (DCP) to sweep and effectively identify regions of simple (Morse-Smale) and chaotic structurally unstable dynamics in the system. Our analysis provides striking new insights into the complex behaviors exhibited by this and similar systems.

## 1 Introduction

Nonlinear wave and mode interactions often result in complex dynamics. Remarkably, already elementary systems of two and three weakly interacting

Krishna Pusuluri
Neuroscience Institute, Georgia State University, Petit Science Center, 100 Piedmont Av., Atlanta, GA 30303, USA, e-mail pusuluri.krishna@gmail.com

Arkady Pikovsky
Department of Physics and Astronomy, Potsdam University, Karl-Liebknecht-Str 24/25, Bld. 28, D-14476 Potsdam, Germany, e-mail pikovsky@uni-potsdam.de

Andrey Shilnikov
Neuroscience Institute, and Department of Mathematics and Statistics, Georgia State University, Petit Science Center, 100 Piedmont Av., Atlanta, GA 30303, USA, e-mail ashilnikov@gsu.edu

waves can demonstrate chaotic behavior [1, 2, 3, 4, 5, 6, 7]. One such simple model is the so-called Rabinovich system, describing wave interaction with complex dynamics in a system of three resonantly coupled waves, comprised of two parametrically excited waves, and another wave that is in synchronism with this pair [8]. It can exhibit the following states while remaining phase locked:

- the trivial static stabilization of parametric instability at low pump fields, which corresponds to a fixed point with zero wave amplitudes in the phase space;
- the static and cyclic stabilizations of parametric instability elimination, corresponding to stable fixed points with non-zero wave amplitudes, and cyclic oscillations of wave amplitudes, respectively;
- eventually, Lorenz like chaotic behavior with a stochastic stabilization region of self-oscillations of wave amplitudes, at higher pump fields.

The physical motivation for the Rabinovich system proposed in [8] is as follows. A whistler wave with wave vector $\mathbf{q}$ and frequency $\omega_q$, propagates along a magnetic field $\mathbf{H}$ in a non-isothermal magnetoactive plasma. This wave parametrically excites a plasma wave $(\mathbf{k}, \omega_k)$ and an ion sound $(\boldsymbol{\kappa}, \omega_\kappa)$, provided that the resonance conditions $\mathbf{k} + \boldsymbol{\kappa} = \mathbf{q}$, $\omega_k + \omega_\kappa = \omega_q$ are fulfilled. These two parametrically excited waves are resonant with the plasma wave $(\mathbf{k_1}, \omega_{k_1})$ where $\mathbf{k_1} = \mathbf{k}\text{-}\boldsymbol{\kappa}$, $\omega_{k_1} = \omega_k$ - $\omega_\kappa$, which is synchronous to the produced pair. As a result, one obtains a closed set of amplitude equations for the three waves $\omega_k, \omega_\kappa$ and $\omega_{k_1}$, where the energy that comes from the constant pump $\omega_q$, is distributed between the waves due to nonlinear resonant coupling, and eventually is dissipated due to linear damping.

The simplified set of equations governing this resonantly coupled wave triplet system – the Rabinovich system – is given by the following equations:

$$
\begin{aligned}
\dot{x} &= hy - \nu_1 x - yz \\
\dot{y} &= hx - \nu_2 y + xz \\
\dot{z} &= -z + xy \,.
\end{aligned}
\tag{1}
$$

Here, $x$, $y$ and $z$ correspond to the amplitudes of the three resonantly coupled waves – the parametrically excited plasma wave $\mathbf{k}$, the parametrically excited ion sound $\boldsymbol{\kappa}$ and the synchronous plasma wave $\mathbf{k_1}$, respectively. Quantities $h, \nu_1$ and $\nu_2$ are the parameters of the system: the value of $h$ is proportional to the pump field, whereas $\nu_1$ and $\nu_2$ are the normalized damping decrements in the parametrically excited waves $\mathbf{k}$ and $\boldsymbol{\kappa}$, respectively. After the original investigation in [8], the studies of this system have further been continued in [9, 10, 11, 12, 13, 14, 15, 16, 17, 18, 19, 20, 21, 22, 23, 24], see also the contribution by S. Kuznetsov in this volume [25].

Although initial numerical simulations have revealed the presence of a Lorenz-like chaotic behavior in the Rabinovich system, the exact boundaries of static, periodic and chaotic dynamics in the parametric space have not been identified. The underlying structures governing the organization of chaos in the system, such as the various homoclinic and heteroclinic connections, and codimension two bifurcation points called the Bykov terminal points (T-points), with characteristic spirals typical for Lorenz-like systems [26, 27, 28, 29, 30, 31], have not been disclosed either. Even as of now, there is only a limited set of computational tools that can be employed to detect such structures in the parametric space of a system, and especially, in Lorenz-like models. In particular, tools based on Lyapunov exponents are computationally effective to sweep, and find regions of stationary (equilibria states), periodic and chaotic dynamics. However, they fail to reveal details of fine or any constructions of homoclinic and heteroclinic structures in the parametric spaces, that are the basic and imperative building blocks of structurally unstable, deterministic chaos in most systems. While parameter continuation techniques let some such structures to be revealed, one has to possess specific skills and enormous patience to perform a painstaking reconstruction of the bifurcation unfolding of the system in question, in its 2D parameter plane, by separately following a few dozens of principal bifurcation curves, one after the other [30, 31].

One of the aims of the current study is to discuss and demonstrate a recent advance in the field that became possible with the development of a suite of computational tools utilizing symbolic representations of simple and chaotic dynamics. This allows for fast and effective identification of bifurcation structures underlying, and governing, deterministic chaos in systems with the Lorenz strange attractors, as well as those with spiral chaos with the Shilnikov saddle-focus [32, 33, 34, 35]. Moreover, the latest advances in GPU and parallel computing techniques have empowered us to achieve a tremendous degree of parallelization to reconstruct bi-parametric sweeps, at a fraction of the time taken for traditional serial computational approaches, for a comparable analysis. In this paper, we employ this computational toolkit to disclose the bifurcation features of complex dynamics in the Rabinovich system.

In the following sections, we will analytically describe, and numerically simulate, the solutions of the Rabinovich system. We will then describe the symbolic apparatus and the computational techniques, and apply them to study this system. Next, we will present our results and identify various important structures that provide a framework to organize the complex dynamics arising in the system. A brief description of the methods used in the study is presented towards the end.

## 2 Solutions of the Rabinovich System

The system (1) is $\mathbf{Z}_2$-symmetric – i.e., invariant under the involution $(x, y, z) \leftrightarrow (-x, -y, z)$. All of its trajectories are confined within an ellipsoid given

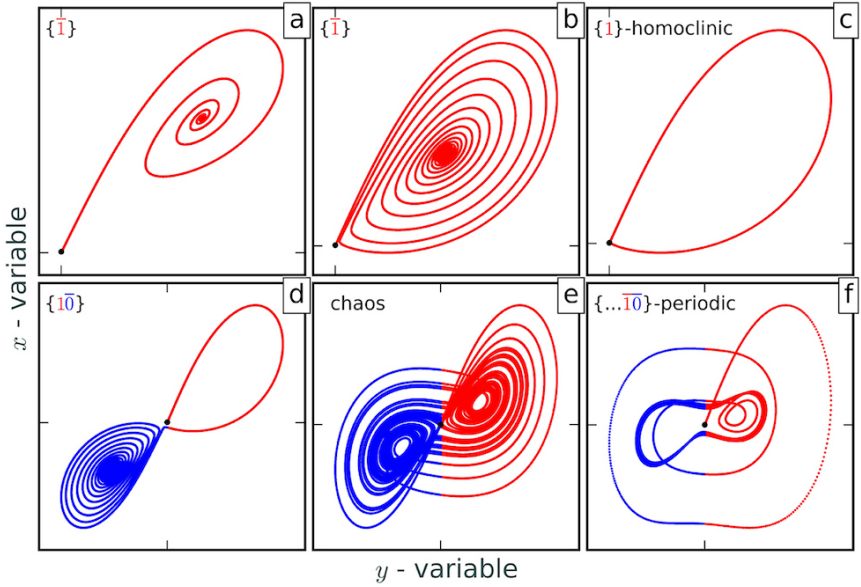

**Fig. 1** Snapshots of the dynamics of the right unstable separatrix $\Gamma_1$ of the saddle $O$ (black dot) in the Rabinovich system, in the $(x, y)$-projection. Sections of $\Gamma_1$ are color-coded in red when $y > 0$, and in blue when $y < 0$, for the sake of clarity. Symbolic representations are as described in Section 3. A bar in the symbolic representation denotes a repetitive sequence. The parameters are set as $\nu_1 = 1, \nu_2 = 4$. At $h = 3$, $\Gamma_1$ converges to the stable focus $C^+$ in **(a)**, and comes close to the saddle $O$ at $h = 3.95$ in **(b)**, generating a persistent sequence $\{111...\overline{1}\}$. After the primary homoclinic orbit of the saddle at the origin at $h \simeq 3.9998$ in **(c)**, $\Gamma_1$ converges to the stable focus $C^-$ at $h = 4.1$ to generate the sequence $\{100...\overline{0}\}$ in **(d)**. Chaotic attractor is seen in the system at $h = 8$ in **(e)**, and convergence to the periodic attractor $\{\overline{10}\}$ at $h = 18$ in **(f)**.

by $u \leq 9h^2k^{-1}$, where

$$u = 2x^2 + y^2 + (z - 3h)^2$$
$$\dot{u} \leq -ku + 9h^2 \qquad (2)$$

At low pump amplitudes $h < (\nu_1\nu_2)^{\frac{1}{2}}$, the system has just a single equilibrium state $O(0,0,0)$, which is a global attractor of the system, pulling all trajectories inwards. In this state, the system is below the threshold of parametric instability. This equilibrium state $O$ undergoes a pitchfork bifurcation at $h = (\nu_1\nu_2)^{\frac{1}{2}}$, that gives rise to two more equilibrium states $C^{\pm}$ $(\pm(z^0l)^{\frac{1}{2}}, \pm(z^0l)^{\frac{1}{2}}, z^0)$ for larger pump fields $h > (\nu_1\nu_2)^{\frac{1}{2}}$, where $z^0 = (h^2 - \nu_1\nu_2)^{\frac{1}{2}}$ and $l = (h - z^0)\nu_1^{-1}$. After the bifurcation, i.e., beyond the parametric instability threshold, the zero equilibrium state $O$ becomes unstable, resulting in parametric instability elimination. The fixed points $C^{\pm}$ can be either stable or unstable, depending upon the parameters of the system $h, \nu_1$ and $\nu_2$. Stable equilibria $C^\pm$ correspond to a static stabilization of parametric instability elimination (Fig. 1a,b and d). Equilibria $C^\pm$ lose stability through an Andronov-Hopf bifurcation that gives rise to a pair of stable periodic orbits, corresponding to stable cyclic self-oscillations of wave amplitudes (Fig. 1f). Besides periodic oscillations, the Rabinovich system may possess a Lorenz-like strange attractor within the finite-size ellipsoid, with stochastic variations of wave amplitudes, which is associated with chaotic saturation of the parametric instability (Fig. 1e). Here, the origin is a saddle with a two-dimensional stable manifold and a pair of one-dimensional unstable separatrices, while the equilibrium states $C^\pm$ are saddle-foci with one-dimensional incoming separatrices and two dimensional unstable manifolds. In addition, like the original Lorenz model, the Rabinovich system can be bi-stable with coexisting stable equilibrium states $C^\pm$ and the strange attractor, see [8]. Also, note that the three resonating waves remain phase locked while their amplitudes exhibit the above mentioned complex behaviors (see [9] for a further exploration of this phase locking).

A homoclinic bifurcation occurs in the system when both outgoing separatrices of the saddle (or of either saddle-focus) come back to it along the 2D stable manifold. Fig. 1c illustrates a single "right" separatrix of the saddle at the homoclinic bifurcation. Before and after the primary homoclinic bifurcation, the separatrix spirals converging towards either $C^+$ or $C^-$ (Fig. 1b,d). Similarly, a one-way heteroclinic connection occurs when the outgoing separatrix of the saddle $O$ connects with either of the saddle-foci $C^\pm$ by merging with a 1D incoming separatrix (see Fig. 6). Note that such connections always come in pairs, due to $\mathbf{Z}_2$-symmetry.

## 3 Symbolic Representation

The fundamental feature of the Lorenz attractor is that it is both dynamically and structurally unstable [36, 37]. A trademark of any Lorenz-like system is the strange attractor of the iconic butterfly shape, as the one shown in Fig. 1e. The "wings" of the butterfly are marked with two symmetric "eyes" containing equilibrium states $C^+$ and $C^-$, stable or not, isolated from the trajectories of the Lorenz attractor. This attractor is structurally unstable as it bifurcates constantly as the parameters are varied. The primary cause of structural and dynamic instability of chaos in the Lorenz equations and similar models is the singularity at the origin – a saddle with two one-dimensional outgoing separatrices. Both separatrices densely fill the two spatially symmetric wings of the Lorenz attractor in the phase space [26]. The Lorenz attractor undergoes a homoclinic bifurcation when the separatrices of the saddle change the alternating pattern of switching between the butterfly wings centered around two other symmetric equilibria, which can be stable foci or saddle-foci de-pending on the parameter values. At such a change, the separatrices come back to the saddle, thereby causing a homoclinic explosion in phase space. The computational approach that we employ for studying Lorenz-like and similar systems, capitalizes on the key property of deterministic chaos – the sensitive dependence of solutions on variations of control parameters. In particular, for the Lorenz-type attractors, chaotic dynamics are characterized by unpredictable flip-flop switching between the two spatial wings of the strange attractor, separated by the saddle singularity at the origin. This is the main reason why the saddle $O$ is the primary source of instability in such systems, including the Rabinovich system. The ideas of this computational research are greatly inspired by, and deeply rooted in, the pioneering studies of L.P. Shilnikov [38, 39, 40]. His extensive knowledge of homoclinic bifurcations helped to transform the theory of strange attractors into a mathematical marvel [26, 41, 42, 43]. The reader may find more detailed information about the Lorenz-like systems and symbolic computations in the original papers [32, 33, 34, 35].

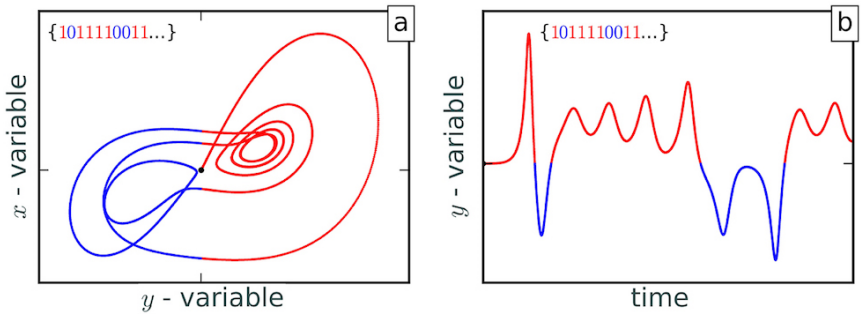

**Fig. 2** Symbolic representation of the right separatrix $\Gamma_1$ at the parameter values $\nu_1 = 1$, $\nu_2 = 4$, and $h = 9$. **(a)** Evolution of the trajectory in the phase space, projected onto the (xy)-plane **(b)** Time evolution of the $y$-variable. The principle of symbolic encoding : each portion of the trajectory turning around the equilibrium state $C^+$ (red section) is represented by 1, whereas each portion looping around $C^-$ (blue section) is represented by 0. The trajectory shown is thus converted into the symbolic sequence {1011110011...}.

In order to identify regions with topologically identical dynamics in the parametric space, we follow the time progression of a single trajectory - the right outgoing separatrix $\Gamma_1$ of the saddle at the origin. We convert the flip-flopping patterns of this trajectory around $C^\pm$ (see Fig.2) into a binary symbolic sequence $\{k_n\}$ obeying the following rule :

$$
k_n = \begin{cases} 1, & \text{when the the separatrix } \Gamma_1 \text{ turns around } C^+, \\ 0, & \text{when the the separatrix } \Gamma_1 \text{ turns around } C^-. \end{cases}
$$

Alternatively, one can detect relevent events of $\frac{dy}{dt} = 0$, provided that $\frac{d^2y}{dt^2}$ is negative or positive for 1 or 0, respectively, see the sampled trace $y(t)$ in Fig. 2b. We use an overbar symbol to represent repetitive sequences. For example, the periodic orbit turning once around $C^+$, then once around $C^-$ and so on, generates an infinite repetitive sequence $\{1010101010...\}$, or $\{\overline{10}\}$ for short.

For a sequence $\{k_n\}$ of length N, starting with some $(j+1)$-th symbol (the very first $j$-transients are skipped), we define a formal power series as follows:

$$P(N) = \sum_{n=j+1}^{j+N} \frac{k_n}{2^{(N+j+1)-n}} \qquad (3)$$

This series is convergent, with its limit ranging between 0 and 1. Whenever the followed separatrix, or any other such trajectory, after some initial transient dynamics, orbits only around $C^+$, so that its $y(t)$-coordinate always remains positive, the corresponding binary symbolic sequence contains only 1s, i.e. $k_n = 1$, and therefore, $P(N) = 1$ in the limit as $N \to \infty$. In the case where the trajectory continuously orbits only around $C^-$ after some transient, so that $y(t) < 0$, $k_n = 0$ and $P(N) = 0$. Otherwise, periodic or aperiodic flip-flopping between and around equilibria $C^+$ and $C^-$, generate either regular or chaotic sequences of 1s and 0s, so that $0 \leq P \leq 1$. This power series provides a way to uniquely quantify the dynamics of the system for a given set of parameters, making it a dynamic invariant. Two different sets of parameters with the same dynamic invariant value, show topologically identical behavior. The way we define this power series is slightly different from how it was previously defined, with the current definition giving increasingly higher weights to symbols towards the end of the sequence, rather than the beginning, see [32, 33, 34]. This lets us achieve a greater contrast in the bi-parametric scans, and thereby, revealing greater dynamical details, between neighboring regions of largely similar dynamics, that differ only in the last symbol in their binary sequences due to homoclinic curves separating such regions (see Section 5 – Methods). Alternatively, one can also convert the binary sequences into a decimal representation.

## 4 Bi-parametric Scans with Symbolic Computations

In order to obtain bi-parametric scans, we keep one of the three parameters of the system – $\nu_1$, $\nu_2$ or $h$, constant, while varying the other two. For each set of parameters in the bifurcation plane, we always follow the positive unstable separatrix $\Gamma_1$ of the saddle at the origin in the Rabinovich system (1). Note that, as the system is $Z_2$-symmetric, our results stay the same even if the left separatrix of the saddle is followed, provided there is the swapping of the symbols $0 \rightleftarrows 1$, resulting in the same symbolic sequence $\{k_n\}$, and the corresponding invariant $P(N)$ values ranging within $[0, 1]$, by the above construction. These invariant values are then projected on to the 2D parametric space, using a colormap that can uniquely identify up to $2^{24}$ different values of $P(N)$, via the whole spectrum of colors. This results in the desired bi-parametric scans, such as the ones sampled in Figs. 3, 5, and 7-10. Regions corresponding to similar dynamics, that generate identical symbolic sequences of a given length, and therefore, carry the same dynamic invariant values $P(N)$, are identified by the same colors in the bi-parametric bifurcation sweep of appropriate resolution.

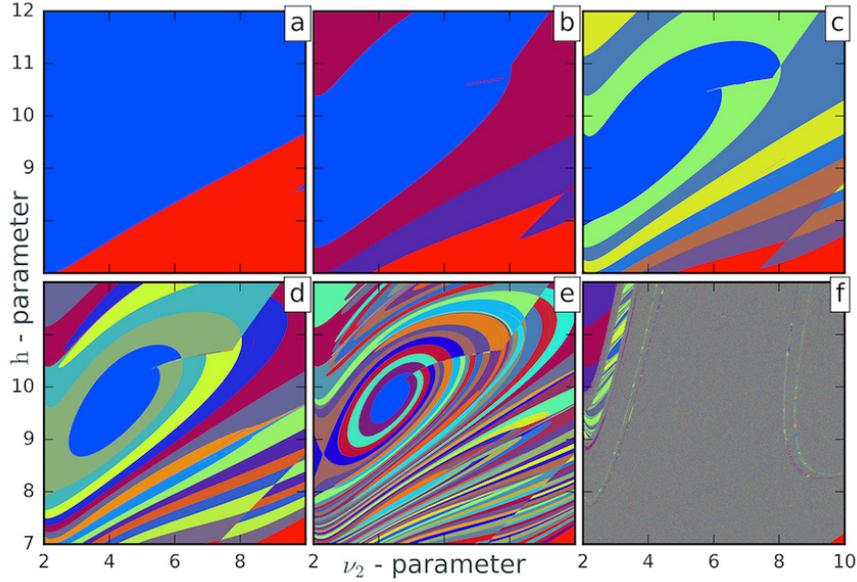

**Fig. 3** Emergent chaos via homoclinic explosion in bi-parametric $(\nu_2, h)$-scans with a fixed $\nu_1 = 1$. **(a)** Bi-parametric scan of length 3, $n : 1 - 3$. The secondary homoclinic curve encoded as {10} (see Fig. 4b) divides the parametric space into two distinct subregions : {100} (red) (see Fig. 4a) and {101} (blue) (see Fig. 4c); **(b)** A longer scan with the first 4 symbols, $n : 1 - 4$, reveals an additional homoclinic curve, {100}, dividing the red region {100} (red) of (a) into two subregions: {1000} and {1001}, while the blue region {101} of (a) is partitioned into two subregions {1010} and {1011} by the homoclinic curve {101}; Sweeps with **(c)** $n : 1 - 5$, **(d)** $n : 1 - 6$ and **(e)** $n : 1 - 8$, gradually disclose finer underlying structures of homoclinic bifurcation unfolding; **(f)** Multi-colored, "noisy" sweep with a long sequence of $n : 105 - 128$, is indicative of a region with structurally unstable, chaotic dynamics in the system (see Section 4.3.1 - Deterministic Chaos Prospector with Periodicity Correction).

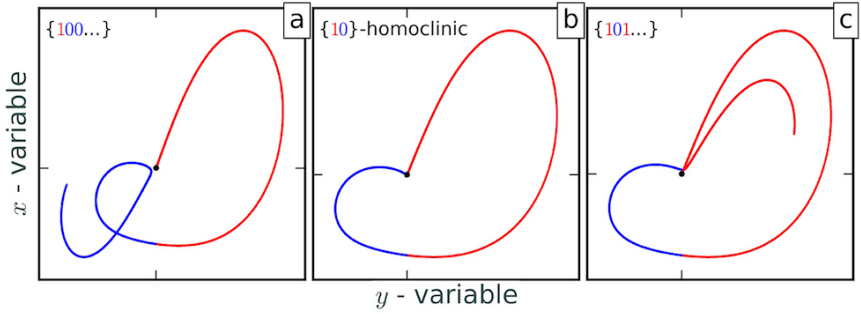

**Fig. 4** Progressive snapshots of the right 1D separatrix of the saddle in different sub-regions of Fig. 3a with a fixed $\nu_1 = 1$, decoded with the first three symbols. **(a)** Dynamical and symbolic representation {100} at $\nu_2 = 4$, $h = 7.5$; **(b)** Secondary homoclinic orbit {10} at $\nu_2 = 4$, $h \simeq 7.6$; **(c)** At $\nu_2 = 4$, $h = 7.7$, the trajectory corresponds to the symbolic sequence {101}.

## 4.1 Emergence of Chaos via Homoclinic Explosion

In this section, we demonstrate how the symbolic computations technique can gradually reveal the complex organization of dynamics, and the underlying non-local bifurcations in the system. Fig. 3 presents a series of $(\nu_2, h)$-bi-parametric sweeps, $\nu_2 \in [2, 10]$ and $h \in [7, 12]$ and at fixed $\nu_1 = 1$, with increasing length/depth of symbolic sequences, from 3 through 8, along with the case of $n : 105 - 128$ range. As such, at every point in the given region, the symbolic dynamics remain identical up to the first two symbols {10}, because the right separatrix always make its first loop around $C^+$ with $y > 0$, followed by a loop around $C^-$ with $y < 0$. As the length of the trajectories, and therefore, of the sequences generated, is increased, initially upto 3 symbols, i.e. $n : 1, 2, 3$, the scan detects a secondary homoclinic bifurcation curve that separates the two sub-regions shown in blue and red in Fig. 3a. In the red sub-region, the separatrix makes the third loop around $C^-$ (Fig .4a), whereas in the blue sub-region, its third loop is around $C^+$ (Fig. 4c). The borderline separating these two sub-regions is a bifurcation curve (arguably of codimension-1), corresponding to the secondary homoclinic orbit of the saddle at the origin (Fig. 4b). This homoclinic bifurcation curve, marked by {10}, divides the parametric space in Fig. 3a into two distinct sub-regions corresponding to the sequences starting with {100...} and {101...}, respectively. Note that both the homoclinic curve and the entire region of Fig. 3a are identified with the same symbolic sequence. As we include one more symbol, i.e., as $n$ runs from 1 to 4 in the computation of $P(N)$, both sub-regions of Fig. 3a, red and blue, are further subdivided by bifurcation curves representing longer homoclinic orbits of the saddle, see Fig. 3b. The homoclinic curve {100} divides the red sub-region {100} of Fig. 3a into two sub-regions coded by $\{1000\}$ and $\{1001\}$ in Fig. 3b. Similarly, the blue sub-region $\{101\}$ in Fig. 3a is subdivided into two sub-regions coded by $\{1010\}$ and $\{1011\}$ in Fig.3b, by the homoclinic bifurcation curve $\{101\}$. Adding more symbols to the computation of the bi-parametric sweep increases its depth, lets us detect more complex homoclinic bifurcations, and gradually reveals the underlying structures that result in the complexity of the system (Fig. 3c,d,e). In this case, the complexity is organized around a central point called a terminal point (T-point) (Fig. 3e), which will be discussed further in the next section. For very long sequences with $n : 105 - 128$ range, the bi-parametric scan indicates that the system continuously undergoes a plethora of homoclinic bifurcations as the parameters are varied, and exhibits structurally unstable dynamics due to these uncontrollable homoclinic explosions (Fig. 3f).

## 4.2 Heteroclinic connections and Bykov T-Points

Fine organization of the structure of the chaotic region with the primary T-point is revealed in greater detail in Fig. 5. It demonstrates the complex universality and self-similarity of characteristic spirals typical for most Lorenz-like systems. Here, the primary T-point is marked $T1$. At this codimension-2 point, the 1D outgoing (unstable) separatrix of the saddle, after the first two loops $\{10\}$ (common to the entire parametric space under consideration here), merges with the 1D incoming (stable) separatrix of the saddle-focus $C^+$, thus forming a one-way heteroclinic connection (Fig. 6a,e). Note that both the saddle and the saddle-focus have stable and unstable manifolds that transversally intersect in the 3D phase space of the Rabinovich system. This makes the heteroclinic connection closed as $t \to \pm\infty$. Thus, at the T-point $T1$, the separatrix makes an infinite number of revolutions around $C^+$ before it comes back to the saddle. As such, its symbolic representation is given by the sequence $\{10\bar{1}\}$, where the overbar represents a repetitive subsequence. As we move away from $T1$ along the adjacent spiral in the parameter space, the number of revolutions of the separatrix around $C^+$ keeps decreasing, and becomes finite. It is known that a Lorenz-like system, near a T-point, exhibits a multiplicity of secondary T-points with increasing complexity, called as Bykov T-points [26, 27, 28, 29, 30, 31, 44]. The short parametric scan in Fig. 5 detects several notable secondary T-points marked as $T2$, $T2'$, $T3$, $T3'$, $T4$ and $T4'$, and the spiral structures associated with them. At $T4$, the outgoing separatrix of the saddle $O$, after the initial two loops $\{10\}$, makes one more loop towards $C^+$ and then merges with the 1D incoming, stable separatrix of the other saddle-focus $C^-$. That is why, this heteroclinic connection is symbolically represented as $\{101\bar{0}\}$ (Fig. 6b,f). Similarly at $T4'$, the outgoing separatrix, after the initial two loops $\{10\}$, makes a loop towards $C^-$, and then hits $C^+$. Thus, this heteroclinic connection is represented as $\{100\bar{1}\}$ (Fig. 6c,g). At $T2$, the complexity of the heteroclinic connection fur-ther increases : after the two initial loops $\{10\}$, the next two loops are around $C^+$, followed by one loop around $C^-$, and then the separatrix comes back to $C^+$; its coding is given by $\{10110\bar{1}\}$ (Fig. 6d,h). The T-points depicted in Fig. 5 can be summarized as follows:

- $T1$: $\{10\ \bar{1}\}$ – Two loops followed by heteroclinic connection to $C^+$
- $T4$: $\{101\ \bar{0}\}$ – Three loops followed by heteroclinic connection to $C^-$
- $T4'$: $\{100\ \bar{1}\}$ – Three loops followed by heteroclinic connection to $C^+$
- $T3$: $\{1011\ \bar{0}\}$ – Four loops followed by heteroclinic connection to $C^-$
- $T3'$: $\{1010\ \bar{1}\}$ – Four loops followed by heteroclinic connection to $C^+$
- $T2$: $\{10110\ \bar{1}\}$ – Five loops followed by heteroclinic connection to $C^+$
- $T2'$: $\{10111\ \bar{0}\}$ – Five loops followed by heteroclinic connection to $C^-$

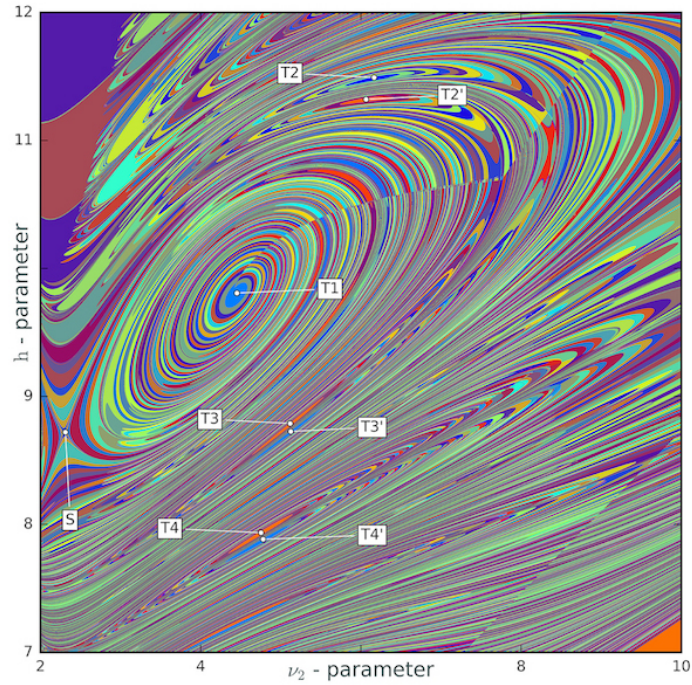

**Fig. 5** Self-similar organization of Bykov T-Points and heteroclinic connections : Bi-parametric $(\nu_2, h)$-sweep at $\nu_1 = 1$, with $n : 5 - 12$ range, detects – primary T-point: $T1 - \{10\bar{1}\}$; secondary T-points : $T2 - \{10110\bar{1}\}$, $T2' - \{10111\bar{0}\}$, $T3 - \{1011\bar{0}\}$, $T3' - \{1010\bar{1}\}$, $T4 - \{101\bar{0}\}$, and $T4' -- \{100\bar{1}\}$. A primary saddle in the parametric space is marked with the symbol $S$. Heteroclinic connections with corresponding $y$-progressions at some of these T-points are presented in Fig. 6.

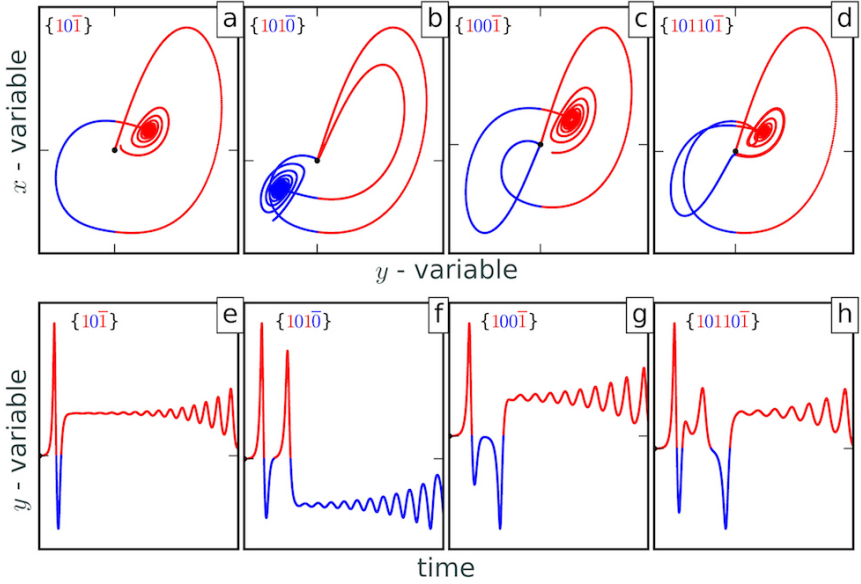

**Fig. 6** T-point configurations **(a)-(d)** and matching time progressions **(e)-(h)** : Heteroclinic connections and the $y$-variable time evolutions at : **(a)**, **(e)** $T1$ – {$10\bar{1}$}; **(b)**,**(f)** $T4$ – {$10\bar{1}\bar{0}$}; **(c)**,**(g)** $T4'$ – {$100\bar{1}$}; and **(d)**,**(h)** $T2$ – {$10110\bar{1}$}, as depicted in the bifurcation diagram in Fig. 5.

## *4.3 Global Bifurcations and Organization of Chaos*

In this section, we study the global organization of chaos using bi-parametric sweeps of the $(\nu_2, h)$-parameter plane (Figs. 7 and 8), and of the $(\nu_1, \nu_2)$-parameter plane (Figs. 9 and 10). We begin this discussion with Fig. 7, showing the $(\nu_2, h)$-sweep with $n : 5 - 12$ range, at $\nu_1 = 1$. It detects several low-order T-point-like structures, labeled by $T1$ through $T7$, as well as characteristic spirals – bifurcation curves of homoclinic orbits and separating saddles. A small sub-region of this diagram nearby $T1$, is magnified in Fig. 5 of Section 4.2. Fig. 8a presents a deeper/longer bi-parametric sweep with $n : 105 - 128$ range. Here, we skip a relatively long initial transient of the separatrix, to reveal the long-term dynamics of the Rabinovich system. The underlying idea here, is a sweep utilizing a typical trajectory of the Rabinovich system, which does not necessarily have to be the separatrix that is employed for the purpose of homoclinic structures. With this new approach, we can reveal the occurrence of chaotic, structurally unstable dynamics emerging through homoclinic explosions, and detect these regions in the parameter space. Such a region in the parameter plane appears to look like a noisy region, due to the interference of multiple colors corresponding to constantly changing $P(N)$, due to homoclinic bifurcations that densely fill in, as the control parameters are varied. On the other hand, regions corresponding to structurally stable (normally hyperbolic) dynamics due to Lyapunov stable equilibria and periodic orbits, i.e., the so-called *Morse-Smale* systems, are coded with solid colors. Note that the same color throughout a region or across regions, corresponds to topologically identical dynamics, by construction.

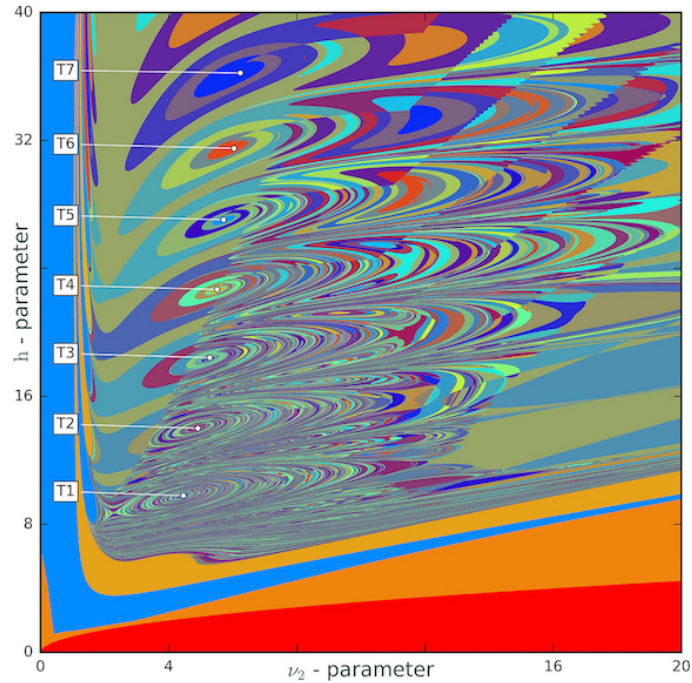

**Fig. 7** Bi-parametric $(\nu_2, h)$-sweep at $\nu_1 = 1$, with $n : 5 - 12$ range, discloses a recursive series of Bykov T-point-like structures marked as T1–T7, saddles, as well as regions of stable periodic dynamics (solid colors) of the system, in the bifurcation diagram.

### 4.3.1 Deterministic Chaos Prospector using Periodicity Correction

A discernible problem of consequence with the symbolic representation of stable periodic orbits of complex configurations existing in the Morse-Smale systems, is their shift-symmetry or shift-circularity feature. For example, the following four sequences: $\{\overline{0110}\}$, $\{\overline{0011}\}$, $\{\overline{1001}\}$ and $\{\overline{1100}\}$, represent the same [stable] periodic orbit, which can be either symmetric or asymmetric in the phase space. To compare whether two such orbits are topologically conjugate or not, at least in their symbolic representation, one has to come up with a consistent rule to sort out and normalize all of their corresponding binary sequences. This issue becomes principal when one deals with skipping arbitrarily long initial transient of typical trajectories converging to a periodic attractor of the system, at the given parameter values.

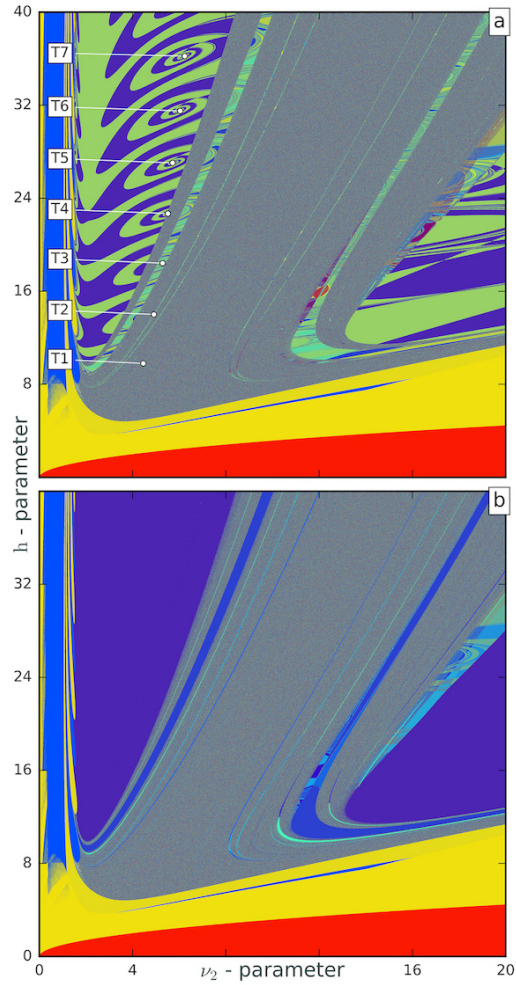

**Fig. 8** Deterministic Chaos Prospector in action : Bi-parametric $(\nu_2, h)$-sweep at $\nu_1 = 1$ with $n : 105 - 128$ to study long-term dynamics of the system : **(a)** Sweep without periodicity correction reveals spiraling artifacts, pseudo T-points, as well as Bykov T-points (labeled by $T1$ to $T7$). **(b)** Sweep enhanced with periodicity correction eliminates spiraling artifacts due to transient dynamics in the existence region of stable periodic orbits. It also detects multiple stability windows (parallel bands of solid colors) representing stable dynamics within the otherwise chaotic regions.

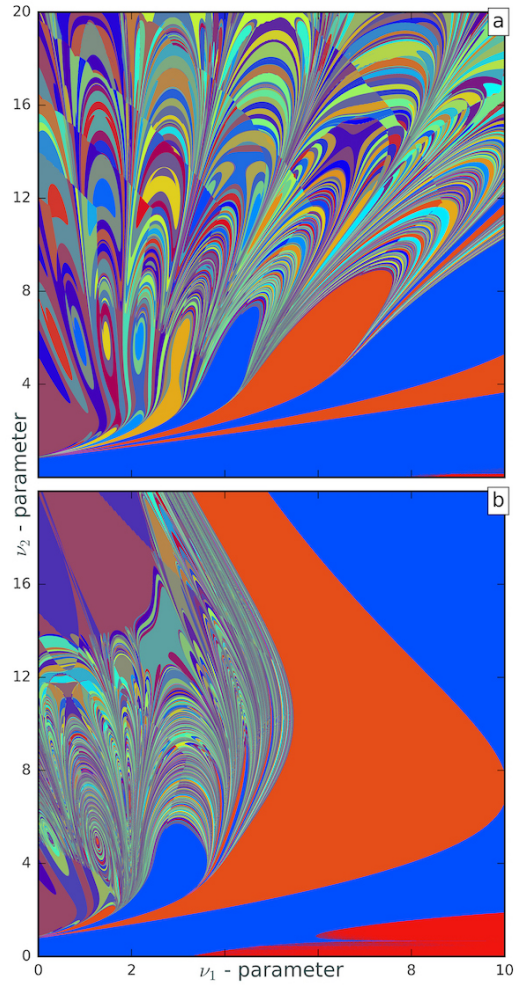

**Fig. 9** Bi-parametric $(\nu_1, \nu_2)$-sweeps with $n : 5 - 12$ range at **(a)** $h = 35$, and **(b)** $h = 15$, reveal universality and organization of the chaos-land, featuring characteristic spirals and saddles that are embedded into the solid-color regions of stable dynamics.

Let us re-iterate : the transient history of two orbits, before they settle down on the same, or topologically same, periodic orbit can be quite different. Consider a transient trajectory converging to a figure-8 periodic orbit that alternatively loops around $C^+$ and $C^-$ back and forth. This orbit admits two symbolic descriptions: $\{\overline{10}\}$ and $\{\overline{01}\}$, which are differently color-mapped on to the parameter space. The existence of such an orbit is detected by the $105 - 128$-range sweep shown in Fig. 8a; here, we skip $104$ initial symbols. Depending on the transient behavior and whether the following 105th symbol is either "0" or "1", the point in the corresponding region is color-coded differently, even though the global attractor is the same figure-8 periodic orbit. As a result, this sweep detects faulty spiral structures, representing $\{\overline{10}\}$ or $\{\overline{01}\}$ sequences, shown in blue and light green colors, respectively, around centers labeled as $T5, T6$ and $T7$ in Fig. 8a, along with some faulty saddles, which happen to be artifacts of the implemented simulation approach.

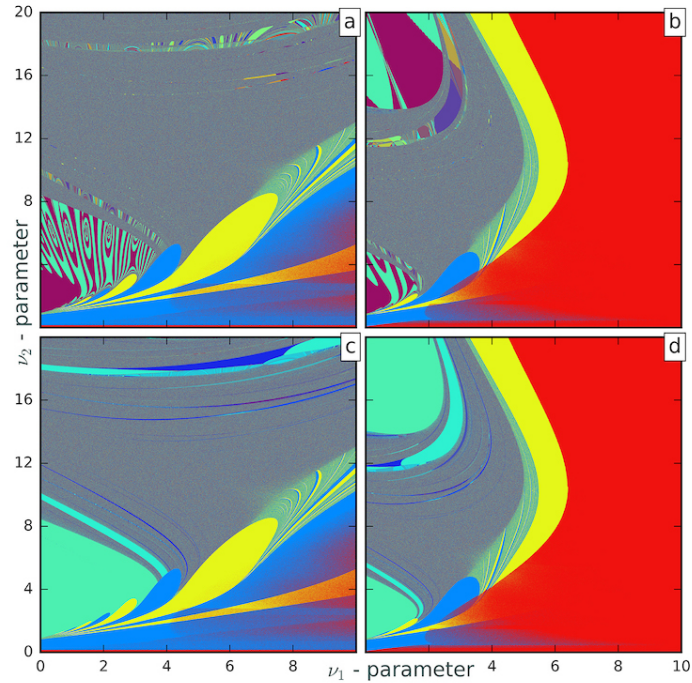

**Fig. 10** Deterministic Chaos Prospector in action : Bi-parametric $(\nu_1, \nu_2)$-sweeps with $n : 105 - 128$ range at $h = 35$ in panels **(a),(c)** and at $h = 15$ in panels **(b),(d)**, detect regions of chaotic, structurally unstable dynamics, and simple, Morse-Smale dynamics due to the existence of stable equilibria and periodic orbits, in the "noisy" and solid-colored regions in the parameter plane, respectively. Artificial spiral structures appearing in **(a),(b)** are eliminated with the periodicity correction technique used in **(c),(d)**. Panels **(c),(d)** reveal multiple stability windows of simple dynamics in the otherwise chaotic – "noisy" regions.

In order to overcome this issue, we developed the technique of "Periodicity Correction" to detect periodic orbits, and to determine their periods using a circular permutation approach. This allows us to consistently choose the same symbolic representations for similar periodic orbits, correctly compute the corresponding $P(N)$-value, and colormap it to the parameter plane. As with the above example, identical periodic sequences $\{\overline{10}\}$ and $\{\overline{01}\}$ are normalized to $\{\overline{01}\}$ to determine its $P$-value. Similarly, the three representations – $\{\overline{101}\}$, $\{\overline{110}\}$ and $\{\overline{011}\}$ for the same (or topologically similar) periodic orbit(s) of periodicity 3, are normalized to the smallest valued binary sequence $\{\overline{011}\}$ to evaluate the $P(N)$-value, and the corresponding color-code (see Section 5 – Methods).

Fig. 8b presents the sweep of the same resolution and depth as Fig. 8a, but using periodicity correction. One can see that, with this technique, the diagram is free of the aforementioned spiral artifacts in the region (dark blue) of existence of the stable figure-8 periodic orbit corresponding to the sequences $\{\overline{10}\}$ or $\{\overline{01}\}$. Nevertheless, the presence of those spirals in the scan without periodicity correction (Fig. 8a) indicates the existence of heteroclinic connections close to the centers of those pseudo T-point spirals - $T5, T6$ and $T7$. Indeed, the points labeled by $T1$ and $T2$ appear to be genuine T-points with quite complex long-term dynamics, whereas, the points labeled by $T3$, $T5$, $T6$ and $T7$ correspond to stable heteroclinic connections between the saddle-foci $C^+$ and $C^-$, that the transitioning figure-8 periodic orbit approaches in the limiting case. The point $T4$ is located next to the boundary between stable and chaotic regions, and it is rather difficult to evaluate its contribution at this resolution. In addition, Fig. 8b reveals multiple parallel bands of constant colors with gradually decreasing widths, within the otherwise noisy regions. This is indicative of the presence of stability windows corresponding to regular, periodic dynamics alternating with chaotic behaviors. Bands of such constant colors are not detected in Fig. 8a, which suggests that, in these regions, though the long-term dynamics are ultimately identical, convergence along distinct paths creates a "masquerading" effect that the developed technique of periodicity correction exposes. Overall, we call this technique "Deterministic Chaos Prospector", since it readily identifies regions of simple (Morse-Smale) and chaotic structurally unstable dynamics in the parametric plane.

Fig. 9 presents the $(\nu_1, \nu_2)$-parametric sweeps of the Rabinovich system for two different values of $h$ : $h = 35$ in panel (a) and $h = 15$ in panel (b); in both cases, a short symbolic scanning is done with $n : 5 - 12$ range. Both sweeps disclose a stunning complexity of the organization of the bifurcation unfolding of the system, with a plethora of Bykov T-points with characteristic spirals, and separating saddles in the chaotic region, that greatly stand in contrast to the Morse-Smale regions of simple and stable dynamics. To conclude, long sweeps with the scanning range $n : 105 - 128$ to expose the long-term dynamics of the Rabinovich system at $h = 35$ and at $h = 15$ are presented in Figs. 10a and 10b, respectively. Both indicate the occurrence of chaotic, structurally unstable dynamics, clearly depicted by the seemingly noisy regions. As seen above, periodicity correction gets rid of some of the spiral structures around the pseudo T-points (Fig. 10c,d). With the enhanced technique, we can also easily identify multiple bands or stability windows cor-responding to periodic attractors, within the chaos-land of the system under consideration.

## 5 Methods

Computations of bi-parametric sweeps are performed on a workstation with Intel Xeon(R) 3.5GHz 12-core CPU and 32 GB RAM, with an NVidia Tesla K40 GPU for parallelization using CUDA. A bi-parametric sweep over a grid of $5000 \times 5000$ mesh points, with a scanning depth $n : 5 - 12$ (Fig. 5) takes about 4.45 seconds, whereas, for long-term dynamics with $n : 105 - 128$ (Fig. 3f), it is about 42 seconds. With the periodicity correction algorithm employed, these numbers increase to approximately 4.68 seconds and 43.7 seconds, respectively. Visualizations of the sweeps and trajectories are done using Python. The colormap is constructed by discretizing the range of computed $P$-values, i.e., [0,1], into $2^{24}$ distinct levels and assigning them RGB (Red-Green-Blue)-color values which are arranged in the following order – increasing values of Blue color from 0 through 1, decreasing values of Red color from 1 through 0, and randomly assigned values of Green color between 0 and 1. As such, $P = 1$ is associated with a bluish color, while $P = 0$ with a reddish color. With this colormap, we can identify distinct topological dynamics of up to 24 symbols long. As seen in Fig. 3, since two neighboring regions differ in the last symbol, the corresponding $P$-value is defined in such a way that the weight of the last symbol is the highest, so that the two neighboring regions fall in either half of the color map range $[0, 1]$, and thus, have the greatest contrast. In regions where the equilibrium state at the origin is stable, and is the only global attractor of the system, the $P$ value is complementarily set to $-0.1$, which is outside the normalized range of all computed $P$-values. Similarly, adjacent regions in the parametric space close to the pitch-fork bifurcation, through which the stable origin becomes a saddle with 2D stable and 1D unstable manifold, and also gives rise to a couple of saddle foci $C^{\pm}$, are assigned a $P$ value $-0.05$, which is also outside the normalized computed range. In both cases, this is done due to monotone convergence to the steady state or slow transients, which do not admit proper partitioning of the phase space of the system to generate symbolic representation of its solutions.

In order to construct sweeps with periodicity correction, we first detect periodic orbits in each sequence. For a sequence of length $N$, we check for periodic orbits of periodicity up to $\frac{N}{2}$, starting from 2. If a periodic orbit is detected, it is normalized to the principal cyclic permutation with the lowest numerical value in the sorted list of all of its cyclic permutations. For example, the periodic orbits $\{\overline{110}\}$, $\{\overline{101}\}$, $\{\overline{011}\}$ are all normalized to $\{\overline{011}\}$, which is then used to fill up the symbolic sequence of length $N$ for the computation of the corresponding $P(N)$-value.

## 6 Conclusions

In this study we have shown that

- Symbolic representation is an effective tool to reveal the bifurcation origins of complex, chaotic dynamics in the Rabinovich system, and in other similar systems.
- Bi-parametric scans disclose fine organizational features of deterministic chaos due to complex, self-similar assemblies of homoclinic and heteroclinic bifurcations with a plethora of accompanying T-points, spiral structures, and separating saddles in the fractal, self-similar regions in the parameter plane, that correspond to complex chaotic dynamics. There is no other current computational technique that can reveal the complexity of chaos in the parameter plane with such stunning clarity and completeness.
- The developed technique of periodicity correction can correctly rids of a variety of artifacts due to shift-cycling in symbolic representation of complex periodic orbits, as well as effectively detects stability windows of regular dynamics that are embedded within the chaotic regions.
- We developed "Deterministic Chaos Prospector" – the paradigm based on the concept of structural and dynamic instability, which presents a novel and highly efficient approach to identify the regions of chaotic and stable dynamics in the parameter space of a system under consideration.
- Massively parallel multi-parametric sweeping based on symbolic representation, using general purpose GPU-computing, presents the new generation, optimal time computational method to study a chaotic system that admits a proper partition of its phase space.

**Acknowledgements** AS acknowledges the financial support from RSF grant 14-41-00044 at the Lobachevsky University of Nizhny Novgorod, as well as NSF BIO-DMS grant IOS-1455527. We thank the acting members of the NEURDS (Neuro Dynamical Systems) lab at GSU for helpful discussions and proof-reading the manuscript. We are very thankful to Sunitha Basodi for her deep insights into the computational aspects of GPU programming, and the Periodicity Correction algorithm. We gratefully acknowledge the support of NVIDIA Corporation with the donation of the Tesla K40 GPU used in this research.